\documentclass[conference]{IEEEtran}
\pdfoutput=1
\IEEEoverridecommandlockouts

\usepackage{verbatim}
\usepackage[style=ieee,
bibstyle=ieee,
backend=biber,
minnames=1,
maxcitenames=2,
maxbibnames=6,
doi=true,
isbn=false,
url=false,
date=year,
]{biblatex}

\AtEveryBibitem{\clearlist{language}} 
\def\BibTeX{{\rm B\kern-.05em{\sc i\kern-.025em b}\kern-.08em
    T\kern-.1667em\lower.7ex\hbox{E}\kern-.125emX}}
\usepackage{caption}

\usepackage{subcaption}
\usepackage{hyperref}
\usepackage{textcomp} 
\usepackage[table]{xcolor}
\usepackage[T1]{fontenc}

\usepackage{csquotes}

\usepackage{svg}

\usepackage{adjustbox}
\usepackage{pifont}
\usepackage{array}
\usepackage{booktabs}
\usepackage{multirow}
\usepackage{tabularx}

\newcolumntype{R}[2]{%
    >{\adjustbox{angle=#1,lap=\width-(#2)}\bgroup}%
    l%
    <{\egroup}%
}

\usepackage{xcolor}
\usepackage{tcolorbox}  
\tcbuselibrary{theorems}
\newtcolorbox{mydefinition}{colback=gray!20, boxrule=0pt, sharp corners}

\definecolor{MarkerColor}{RGB}{241, 242, 162}

\newcommand{\linebreakand}{%
  \end{@IEEEauthorhalign}
  \hfill\mbox{}\par
  \mbox{}\hfill\begin{@IEEEauthorhalign}
}
\usepackage{hologo}
\usepackage{listings}
\begin{document}

\twocolumn[
\begin{@twocolumnfalse}
\Huge {IEEE copyright notice}

\vspace{1ex}
\large{\copyright\ 2026 IEEE. Personal use of this material is permitted. Permission from IEEE must be obtained for all other uses, in any current or future media, including reprinting/republishing this material for advertising or promotional purposes, creating new collective works, for resale or redistribution to servers or lists, or reuse of any copyrighted component of this work in other works.}

\vspace{1ex}
{\Large Accepted to be published in the \emph{2026 IEEE Intelligent Vehicles Symposium (IV)}}

\vspace{1ex}
{\large Cite as:}

\vspace{0.2cm}
\noindent\fbox{%
    \parbox{\linewidth}{%
        L.~J.~Brettin, T.~Schr\"ader, K.~Kuhlmann, V.~Schmidt, and M.~Maurer, ``A Companion App for an Autonomous Family Vehicle: Identification of Values for an Autonomous Mobility System,'' in \emph{2026 IEEE Intelligent Vehicles Symposium (IV)}, to be published.%
    }%
}

\vspace{2cm}
\end{@twocolumnfalse}
]

\noindent\begin{minipage}{\textwidth}
\hologo{BibTeX}:
\footnotesize
\begin{lstlisting}[frame=single]
@inproceedings{brettin_companion_2026,
  author={{Brettin}, Leon Johann and {Schraeder}, Tobias and {Kuhlmann}, Kerstin and {Schmidt}, Vanessa and {Maurer}, Markus},
  booktitle={2026 IEEE Intelligent Vehicles Symposium (IV)},
  title={A Companion App for an Autonomous Family Vehicle: Identification of Values for an Autonomous Mobility System},
  address={Detroit, USA},
  year={2026},
  publisher={IEEE. to be published}
}
\end{lstlisting}
\end{minipage}

\title{A Companion App for an Autonomous Family Vehicle: Identification of Values for an Autonomous Mobility System
\thanks{
  This research was carried out as part of the UNICAR\textit{agil} project (FKZ16EMO0285) and the autotech.\textit{agil} project (FKZ01IS22088R). We would like to thank the Federal Ministry of Education and Research (BMBF) for its financial support of the project and all members of the consortium for their contribution to this publication.
}
}
\author{\IEEEauthorblockN{
    Leon Johann Brettin\IEEEauthorrefmark{1},
    Tobias Schräder\IEEEauthorrefmark{1},
    Kerstin Kuhlmann\IEEEauthorrefmark{2},
    Vanessa Schmidt\IEEEauthorrefmark{2},
    and Markus Maurer\IEEEauthorrefmark{1}
}\\ 
\IEEEauthorblockA{
    \textit{TU Braunschweig} \\
    \IEEEauthorrefmark{1}\textit{Institute of Control Engineering}, Braunschweig, Germany\\
    \{l.brettin, t.schraeder, markus.maurer\}@tu-braunschweig.de
}
\IEEEauthorblockA{
    \IEEEauthorrefmark{2}\textit{Department of Traffic and Engineering Psychology}, Braunschweig, Germany\\
    \{kerstin.kuhlmann, vanessa.schmidt1\}@tu-braunschweig.de
}
}

\maketitle

\begin{abstract}

In this paper, we present a companion app for an autonomous vehicle aimed at user groups who would normally require an accompanying person to drive them.
Two aspects of a companion app are presented in this paper:
First, the possibility for a trusted person to track the ride of the  person in need of support and second, to put the settings of the vehicle for persons in need of support in the hands of a trusted person.
In addition, this article describes the requirements and addressed values and discusses the safety-relevant aspects of such a companion app.
We also discuss and identify the values that influence passengers and trusted persons using the companion app.
Overall, a companion app can provide new perspectives and opportunities for people in need of support, allowing them to take advantage of the features offered by autonomous vehicles.
It enables trusted individuals to configure the vehicle according to the passengers needs.
Also such an app can be a mechanism to involve trusted persons in the options given by the vehicle and give them the possibility to adapt the vehicle to the needs of the person in need of support.

\end{abstract}


\section{Introduction}

\begin{figure}[ht]
     \centering
     \begin{subfigure}[b]{0.49\textwidth}
         \centering
         \includegraphics[width=\textwidth]{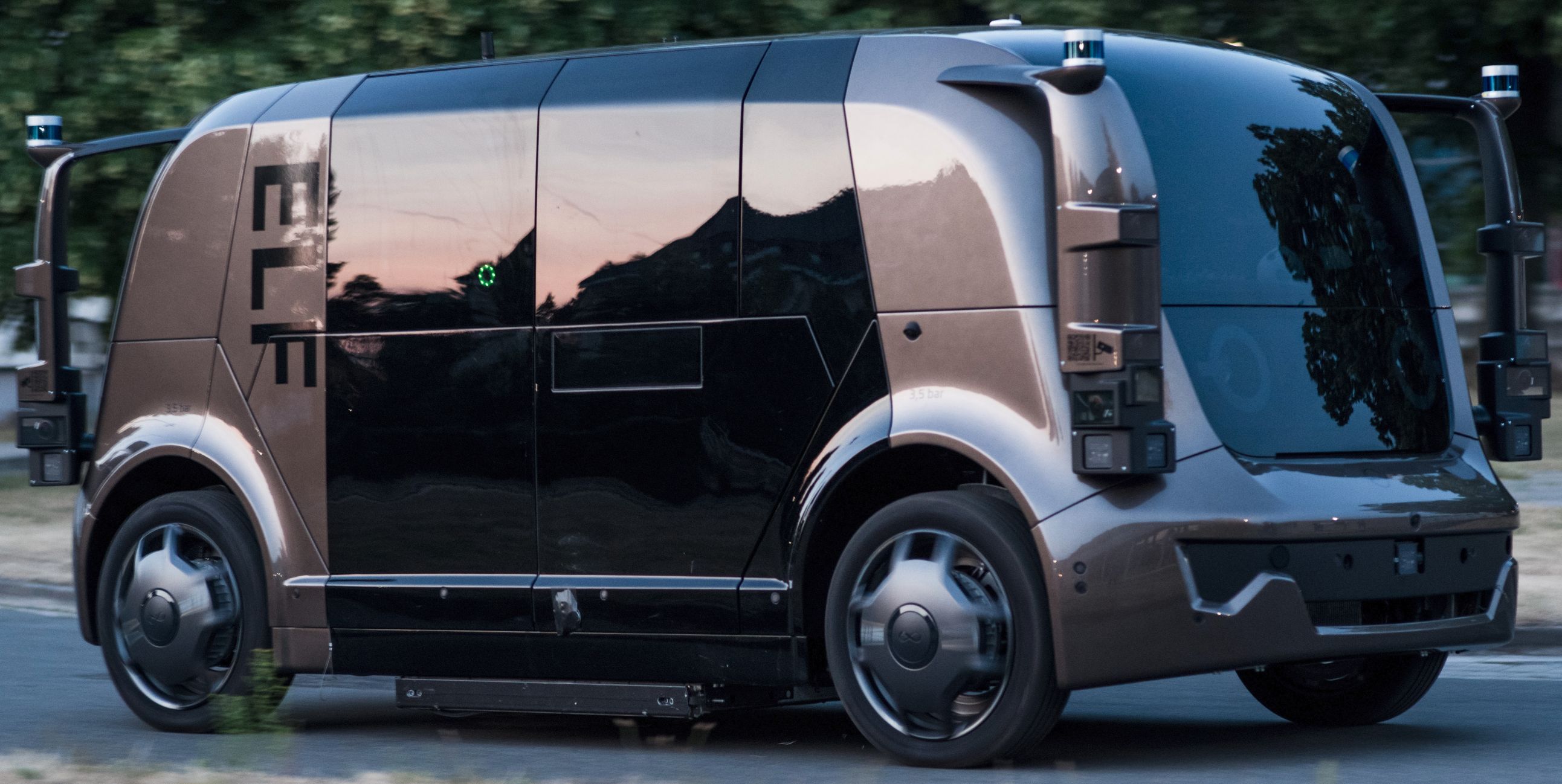}
     \end{subfigure}
     \hfill
     \begin{subfigure}[b]{0.49\textwidth}
         \centering
         \includegraphics[width=\textwidth]{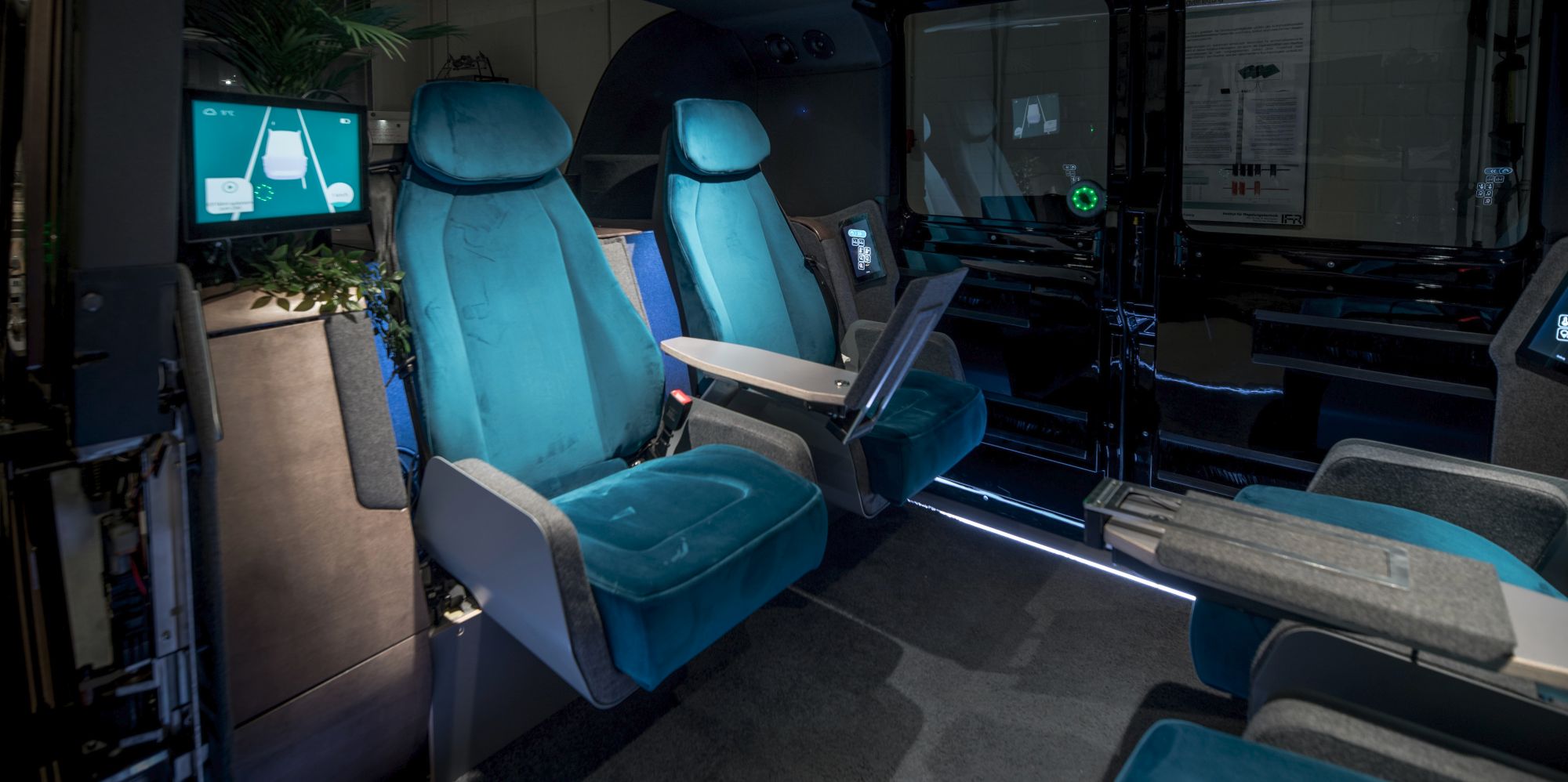}
     \end{subfigure}
        \caption{Pictures of the current vehicle for which the app was developed. The image of the vehicle interior is also used in~\cite{Kuhlmann2025}.}
        \label{fig: elf}
\end{figure}

With the help of autonomous vehicle technologies, people who would otherwise not be able to drive alone, either because they have cognitive or mobility impairments not allowing them to solve problems on their own, because they are too young, have age-related limitations, or are unable to drive for other reasons, will also be able to drive a vehicle independently at some point. 
Autonomous vehicles will therefore expand the user group of drivers to include people in need of protection and support, as the autonomy of the vehicle will enable them to ride alone in such a vehicle even without a trusted person.
The associated rethinking of driving without a person able to drive a vehicle should be an important point in the development process for autonomous vehicles, as this is a way to identify problems with regard to this user group early. 

We developed a smartphone app to help such user groups use an autonomous vehicle.
It gives a trusted person the possibility to adapt the vehicle for the person in need of support, and also gives them the chance to oversee the vehicle from a distance.
The developed app can be separated in two main aspects:
On the one hand, the ability to manage individual settings for each passenger, and on the other hand, the control options for monitoring the passenger in need of support.
Both aspects are presented in this paper.

The research question to which this work contributes is how the interaction with an autonomous vehicle can be improved or in some cases even made possible for the user group in need of support by using the technology of a smartphone app for autonomous driving and what other possibilities, but also problems, can arise from the use of such a companion app.

In a preliminary work for this paper, various user groups with requirements for an autonomous vehicle have already been described \cite{Schraeder2019}.
Motivated by these requirements, a need for such an app as an companion for this user group was recognized.
To demonstrate this approach to the relevant user groups, we developed the app as an Android app and used it in the UNICAR\textit{agil} project for an autonomous prototype vehicle named \textit{auto}ELF, which can be seen in Figure \ref{fig: elf}.

The initial specifications for this app were formulated very openly.
Therefore, an experimental, human centered design approach was used to expand and adapt the companion app.
The results obtained in this work during the development period are based on the values we identified as relevant for the related user groups.
A \textit{value} for this work can be seen as something that a person considers important in life (\cite{Friedman2013} following \cite{Simpson1983}).
The development with a focus on such values is also referred to as \textit{value-based development} and is also applied in research on autonomous driving (see e.g., \mbox{\cite{Friedman2013, Graubohm2020, Nolte2024}}).

This article is structured as follows:
In section \ref{sec: related work}, related work is described.
In section \ref{sec: requirements}, the main aspects of the app are described and in section \ref{sec: values}, the values that the app addresses are presented.
In section \ref{sec: projektkontext}, the project context and the vehicle for which the companion app was developed is shown.
In section \ref{sec: example}, the developed companion app is presented.
Lastly, in section \ref{sec: conclusion}, a conclusion is drawn.

\section{Related Work}
\label{sec: related work}

This section addresses three topics: the state of science and technology, car manufacturing, and other personalization possibilities.

\subsection{Related Work in the Scientific Community}

The use of software tools to manage aspects of a users live is getting more and more common.
Studies and paper like~\cite{Farage2012, Eatock2017, Vespa2018, StatistischesBundesamt2021} show that people are living longer. 
With the help of smartphones, assistive technologies are being developed that enable users to better cope with previously difficult problems.
In some areas of research, the use of apps is already much more popular than in other areas.
Especially in the area of health and smart home, there is already some more research that has been developed with regard to people in need of support:
For example, in the health sector there is research on apps to help people with diabetes keep track of their blood sugar levels \cite{Keith2014}, apps to help blind people navigate in the city~\cite{Ahmetovic2016}, or apps that help people with restricted action planning to be reintegrated into everyday life \cite{Gabel2022}.

In the smart home sector there is research into apps that enable people with disabilities to use voice commands to make things in their home more accessible and enable them to operate appliances and lights \cite{Isyanto2020} or into apps that track older people's daily tasks to help them get through the day and provide caregivers with updates about them \cite{Fahim2012}.

For smartphone apps in combination with autonomous \mbox{vehicles}, \textcite{Ayoub2020} describes a study with a Wizard of Oz vehicle. In this study, children used a wireframe prototype of an app that allowed users to click from one image to another, imitating real app usage for both parents and children. The study concluded that children need to be entertained and buckled in their seatbelts, and parents should be able to track their location and monitor them.
In vehicles, on the other hand, there is other than this study, according to our literature research and to the best of our knowledge, rather less research directed at the development of smartphone apps in combination with a focus on persons in need of support for road vehicles.
This is probably due to the fact that a vehicle is still seen as a means of transport for people who can drive themselves and thus there is a person inside the vehicle who can take care of the person in need of support.
But this is not necessarily the case for an autonomous vehicle.

There are already some approaches in the scientific community that deal with the personalization of driving behavior and the adaptation of the vehicle to specific user group (e.g., \cite{Tulusan2012, Busold2013, Aloul2015}).
But these approaches focus mainly on the driving behavior of the vehicle and not on the combination with a smartphone app that can be used to adapt the vehicle to the needs of the respective user.
\textcite{Trende2019} for example have developed different driving modes that allow the passenger to adapt the driving behavior to what is best suited for the respective user.
\textcite{Kuderer2015} made an attempt to learn a model that adapts to the driver's driving behavior.
In \textcite{Moeller2021}, the temperature behavior in the cockpit of the vehicle was changed so that it is most comfortable for the respective user.
These settings each specify a part of the possibilities that the vehicle has to improve the user experience for the current passenger.

The approaches that come closest to this work are research on apps that develop a virtual assistant for the car:
However, these focus more on making the vehicle itself more tangible and less on managing the possibilities such a vehicle has for a person in need.
An example of this is the research on entertainment for children called PANDA, which is a virtual assistant for children in cars \cite{Gordon2015} and AIDA, a virtual assistant designed to take care of the driver's stress and act as a kind of companion \cite{Williams2013}.

\subsection{Related Work in the Car Manufacturers Community}

Among traditional level 1-3 manufacturers, there is also a wide range of apps for their vehicles.
Each manufacturer has developed its own app so that its customers can use the possibilities of digitalisation in their vehicle.
Most of these apps have a similar basic functionality.
For example, the battery status and consumption can be displayed; the apps keep a history of journeys made; there is some information about the manufacturer displayed and it is possible to use basic comfort functions.
These functions include capabilities like opening and closing the vehicle, changing the temperature in the vehicle, and sometimes opening and closing windows or trunk lids.
What is apparent when looking at the apps is that they mainly focus on these named basic functions of the vehicle. 
Which makes sense, since the apps are designed for this very purpose.

The target group of the app is not, as in this paper, on \mbox{different} user groups, which includes older people and children, but on the owner and driver of the vehicle.
It should be mentioned here, however, that in some vehicles, there are already different driver profiles, which store the driver's settings in the vehicle.
In the Volvo app, for example, the settings are  divided for different users so that there are function settings, such as driver assistance, climate settings, seat settings, etc., and audio and media settings, including settings for navigation, radio and apps \cite{Volvo2021}.

In addition, of course, no functions are offered here that can only be used in fully autonomous vehicles.
However, it is also clear that no value was placed on autonomous driving functions in these apps, since the vehicles do not drive autonomously.

\subsection{Other Ways for Personalization in the Driving Context}

A notable contribution to personalized driving is the integration of Android Auto and Apple CarPlay.
This allows users to integrate their smartphone into the media system of the vehicle.
In this case, it is also possible for each smartphone owner to have their own profile included, as each user has their preferred settings on their own device.
However, the problem here is that the functionality is limited to controls that do not correspond to the driving function, as in the virtual assistant mentioned before (\cite{Gordon2015, Williams2013}).

Another contribution to personalization is provided by apps from ride-hailing services such as Uber, Moia or Waymo.
These providers offer services to book vehicles, to then be driven from one place to the next, just like a taxi.
With Uber, for example, trusted people can be added who can follow the journey via the app~\cite{Uber2022}.

On the Moia app, options can be scheduled in the app in the accessibility section for more time for walking~\cite{MOIA2024_Gehen} and the option for the passenger to have impaired vision so that the driver of the vehicle actively looks for the passenger~\cite{MOIA2024_Sehbehinderung}.

At Waymo, it is already possible to book a driver without a safety driver~\cite{Waymo2024_Safety_Impact}, and more time for walking can be set.
There is also \textit{Assistive in-car audio}, which describes what is happening and where to find the controls while driving.
In addition, it is also possible with Waymo to call special vehicles in which passengers can also get in with a wheelchair~\cite{Waymo2024_Accessability}.
Recently, Waymo developed a program called the ``Waymo Teen Account,'' which allows children aged between 14 to 17 to drive alone with the Waymo vehicles, when a parent gives their permission for such an account.\footnote{See \url{https://waymo.com/teens/} (last accessed on 27.08.2025)}
Here parents have the option to view the progress of the ride on their own smartphone.
Parents have the option of viewing their child's journey progress on their own smartphone.
This is a very recent Waymo development, making it the implementation that most closely resembles the app described in this article.

Because of language differences, it was not possible to investigate ride-hailing services in China in this paper.
These services are usually only available in China and are often offered exclusively in Chinese.\footnote{See e.g., \url{https://apps.apple.com/cn/app/ponypilot/id1488620923} (last \mbox{accessed} on 04.06.2025)}

\vspace{\baselineskip}

In summary, there are already approaches that deal with the personalization of user profiles.
In the scientific community, there is a lot of research in the area of health and home automation and in the area of adapting driving behavior.
In the area of manufacturers, there are approaches, which, however, are mostly more of a kind of toolkit for the driver.  
What is also noticeable, however, is that there is already a move in the direction of more personalization. 
Personalization approaches are most likely to be found in ride-hailing services, as these are also heading in a direction that comes closest to fully automated driving services.
But there is still a need for improvement (for example demanded by \cite{Dorynek2022} for persons with mobility impairments).

\section{Requirements for the Companion App}
\label{sec: requirements}

\begin{figure*}[ht]
  \centering
  \includegraphics[width=0.99\textwidth]{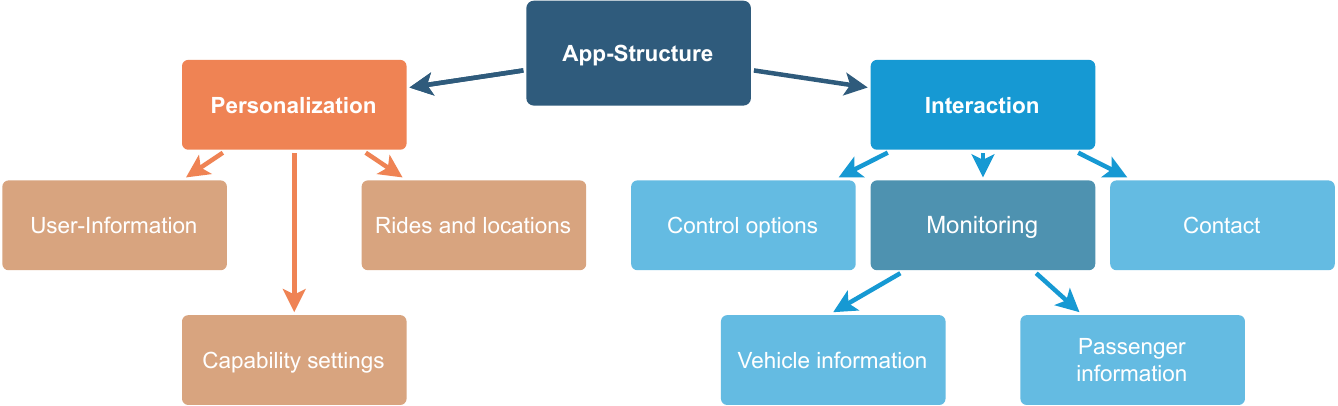}
  \caption{
  Requirements for the companion app.
  The app can generally be divided into two areas: interaction possibilities and personalization possibilities.
  }
  \label{fig:categories} 
\end{figure*}

Based on the project context (see section \ref{sec: projektkontext}), previously conducted requirements analysis methods \cite{Graubohm2020, Schraeder2019} and conducted user studies \cite{Stange2024}, requirements for a companion app were identified.
These requirements can be divided into two aspects:
On the one hand, the direct interaction and monitoring of the vehicle and, on the other hand, the possibilities for personalizing the possible settings of the vehicle.
Both aspects can be further divided into categories, which are shown in Figure \ref{fig:categories} and will be discussed in more detail below.

It should be noted that this is not an exhaustive list of all possible requirements.
This is a collection of requirements that have arisen for this context.
In future work, further requirements may arise or may be deemed invalid.
What requirements are implemented in the app and how they are implemented is shown in section \ref{sec: example}.
These requirements will address different values, which we will present in section \ref{sec: values}.

\begin{itemize}
    \item \textbf{Personalization}
    \begin{itemize}
        \item In \textit{User information}, information about the \mbox{respective} passenger can be stored, such as the type of \mbox{passenger} (e.g., ``child,'' ``adult'' or ``older adults'').
        \item \textit{Capability settings} are user-specific settings, such as settings for visual or hearing impairments or the use of a ramp when getting in and out.
        \item \textit{Rides and locations} refers to settings such as regular rides or restrictions to certain locations.
    \end{itemize}

    \item \textbf{Interaction} 
    \begin{itemize}
        \item \textit{Control options} are interactions such as calling the vehicle or comfort settings.
        \item \textit{Monitoring} describes the ability to access vehicle information or information about passengers in the app.
        \item \textit{Contact} allows trusted persons to contact the \mbox{passengers}. In addition, the app can be used as a contact option for passengers to trusted persons.
    \end{itemize}

\end{itemize}

\subsection{Personalization}
The personalization options allow the user to adapt and set up the vehicle in such a way that it is best suited for the respective passenger.
The focus here is on the fact that the vehicle itself has many possibilities, but many users cannot use them because they do not know or cannot grasp all the possibilities.
For the development of autonomous vehicles this is an important point, because it is no longer just people sitting in the car who can be assumed to have a certain understanding and capability of the vehicle they want to drive, but also people who do not have this skillset or cognitive capabilities in using the vehicle alone.

Older adults who are no longer able to drive, children or people with disabilities should be given the opportunity to use the vehicle alone to a desired destination.
But for every user group other settings may be more important than others.

Some participants of these user groups can be characterized as being either incapable or prohibited from making independent decisions.
For example, a child should not be able to decide completely freely where an autonomous vehicle should go.
On the other hand, an older adult may not be able to fully understand the complexity of setting up a ride to a desired location.

Some settings may also be important for controlling the vehicle.
For example, the prototype vehicle \textit{auto}ELF has a lift that can be used to access the vehicle, and there are settings that enable voice commands in the app for visually impaired passengers.
These adjustments should be made by a person who is capable of doing so and who knows which adjustment options are available and which of them are relevant for the respective driver.
The app should show the individual options for this person in a structured way and thus make tailor-made adjustments for the person in need of support known to him or her.

\subsection{Interaction}

Another important aspect is the ability to look after the occupants of the car remotely.
The app should be able to display the location of the respective passenger and possibly provide further information about the vehicle interior.

For that reason, a companion app can be a means of communication:
It is on the smartphone of the trusted person and thus always available to this person.
This information flow can be seen as a fallback and control level, which potentially increases the safety feeling and trust of those in need of protection in the vehicle.

A certain degree of remote control over the vehicle is even an acceptance criterion for parents to let their children ride in an autonomous vehicle in the first place \cite{Tremoulet2020, Stange2024}.
That control and monitoring can also have negative aspects is discussed in the next section.

\section{Values Addressed by the Companion App}
\label{sec: values}

Based on \textcite{Friedman2013}, \textcite{Graubohm2020} \mbox{developed} values for the most important stakeholder groups of an autonomous vehicle.
The groups are \textit{passengers, secondary users, developers, maintainers}, and \textit{operators}.
As mentioned in the introduction values can be seen as something that a person considers important in life \cite{Friedman2013} (after \cite{Simpson1983}).

This paper focuses on the values of the \textit{passengers} and \textit{secondary users} of the vehicle.
We believe the following values are relevant for the companion app: \textit{Safety}, \textit{Mobility}, \textit{Freedom}, \textit{Security}, \textit{Aesthetics}, \textit{Haptics}, \textit{Efficiency}, \textit{Trust}, \textit{Privacy}, \textit{Sustainability}, \textit{Accessibility} and \textit{Ownership}. 

This section covers the values that a companion app can potentially address.
No claim to completeness is made here, but it is shown which potential possibilities a companion app can have in terms of addressing values.
Also due to limited space requirements, not all values are discussed here.

\subsection{Safety}
The influences on the value of \textit{Safety} are twofold in the case of a companion app, but cannot be compared to concrete safety functions such as an emergency braking function.
With regard to autonomous driving, the companion app can be used to establish an additional channel to trusted persons in an emergency and to establish contact options.
Although this could potentially improve safety, it is a relatively minor consideration compared to the driving function.

The companion app also offers the option of setting \mbox{parameters} that could potentially influence the safety of a ride.
It is important for passengers that these settings are made by a trusted person, so that the system has the necessary information to deal with safety-relevant situations.
The following examples illustrate information that can be set via a companion app and has the potential to influence safety settings:

\begin{itemize}
    \item With the \textit{location restriction} setting, restrictions can be made that limit the ride to certain locations.
     For example, it may be advisable to only allow children to drive to places selected by their parents, in order to avoid potentially dangerous areas.
    \item With the help of \textit{exit conditions}, it is possible to specify that a door is only opened for a child when a known person is waiting for them.
    \item In the case of \textit{simplified operation}, it may be useful for older people, for example, that not every place but only a few places known to the passenger can be selected in order to avoid incorrect entries that send the passenger to completely unfamiliar and thus potentially dangerous places.
\end{itemize}

It is particularly important for passengers who are unable to make their own decisions to have a trusted person who can adjust these settings.

\vspace{\baselineskip}
\subsubsection*{Safety Concerns}
As helpful as the companion app can be, an improper use can also pose a potential danger.
In Nielsen's definition of \textit{usability}, the error type of undetected errors is introduced \cite{Nielsen1993}.
Such errors can be caused in different ways: The settings may have been misunderstood, there may be a typo, or something in the environment of the passengers may have changed after some time of use.
These are also of particular interest in the development of such a companion app, as they can potentially have safety-critical effects on the operation of the vehicle.
The standard ISO 21448 that addresses safety of the intended functionality (SOTIF) should be mentioned at this point, in which the term \textit{reasonable foreseeable misuse} is addressed~\cite{ISO21448}, which describes a similar problem, but is also applicable to other problems than just the user interface.
An example of such an error could look like this:

\begin{itemize}
    \item The child is supposed to be taken to the swimming hall this morning instead of directly to school, because today, exceptionally, swimming is on the schedule in the first period.
    However, the parents have made a typo when specifying the destination of the swimming hall and entered the wrong street.
    The child is now driven to a completely different district and gets out there.
    The vehicle then drives back, as the mission has been completed, and the child is left alone at this location.
\end{itemize}

Of course, there are measures that can be taken to reduce such errors, but it must be pointed out that they are possible.

\subsection{Mobility and Efficiency}

Involving a trusted person can in some cases even enable the use of an autonomous vehicle.
The companion app can be understood as an interface for the trusted person, which simplifies the involvement of this person and thus indirectly enables the \textit{mobility} of the person to be protected.
This can also increase the \textit{efficiency} of the \textit{secondary users}, in this case the trusted person, as they no longer have to ride along as a companion.
However, it should be noted that this is an indirect addressing of this value, as the \textit{efficiency} and \textit{mobility} increase is primarily achieved by the autonomous vehicle and not by the companion app itself.

\subsection{Security and Privacy}

For the value of \textit{Security}, the companion app may have a rather negative effect, as it represents another external access point to the vehicle.
This means that another attack point is created on the vehicle that must be secured.

Another significant aspect is the protection of \textit{privacy}.
The potential surveillance options enabled by the app represent a strong intrusion into the privacy of the passengers.
However, a user study conducted in the vehicle suggests that some people demand passenger surveillance using cameras~\cite{Stange2024}.

Whether this intrusion into privacy is permissible can be discussed at this point.
Here, the value-conflict between the supervisory duty of the trusted person versus the right to privacy of the passengers is reflected (discussions on this can also be found in \cite{Graubohm2020, Stange2024}).

Also the possibility of surveillance and control over the vehicle offers several attack angles for potential attackers on the vehicle.
The information that can be obtained from the vehicle in this way is, to some extent, highly sensitive:
To name just a few examples, regular journeys of the passengers can be queried in the app;
the attacker can find out when which person is driving with the vehicle right now, where they live and where they regularly drive to;
and information about the health status and family relationships can also be uncovered.
In addition, the attacker could change destinations in the vehicle to send the protected passengers to arbitrary places.

All these points must be considered in the development of an overall system.
As is usual with problems regarding risk (e.g., shown in \cite{Fischhoff1978}), a trade-off must be made between the accepted risk and the potential damage caused by the technology, which in this case facilitates or enables autonomous driving for the protected persons.

\subsection{Accessibility}

For the value of \textit{accessibility}, the first aspect is the control options of the app.
Functions that are already available in the apps of the car manufacturers today, such as opening the doors or setting the climate control before getting in, can be counted as \textit{accessibility}.
Also specific aspects of autonomous driving can be found in the companion app.

In the companion app presented in section \ref{sec: example}, one of the most prominent buttons is the request of the vehicle (see Figure~\ref{fig:app details 5}):
Due to the autonomy of the vehicle and the associated possibility of carrying out several journeys of different family members, the vehicle does not have to be parked in front of the owner's house, but can also park somewhere else.
The companion app offers the possibility to operate the vehicle even if it is not on site.

The monitoring functionality allows vehicle information to be retrieved, such as the vehicle's current location, the route it is taking and its battery status.
In addition, passenger information, i.e. who is currently driving with the vehicle and what the person in the vehicle is doing, can be summarized.
These information give the trusted persons a possibility to get an overview of the vehicle and the respective person in need of support.
Furthermore, the \textit{accessibility} to the vehicle is increased, as information about the vehicle and the passengers can be passed on to the trusted person.
This can be particularly useful for parents with children, as it allows a certain degree of control over a new technology such as autonomous driving to be given to the parents.

With the help of settings regarding simplified usability, a companion app addresses the value of \textit{accessibility} for user groups in need of support, as the operation can be adapted to the respective passengers.
Also, such a setting can also enable the journey in the first place, which can be significant for \textit{accessibility}.

Another aspect is the possibility for the trusted person to communicate with the passengers.
In \textcite{Stange2024}, for example, it is mentioned that parents want the possibility to tell the child to sit on the seat or behave in the vehicle, which corresponds to the aspect presented here.

\textcite{Farage2012} for example, describes guidelines for accommodating older adults.
The source mentions options for visual, audio, tactile, mobility, and cognitive accommodations. 
Accessibility can be increased for older adults by setting these options in a way that is tailored to them, making the vehicle easier to use.

One result of the study by \textcite{Kuhlmann2025} was that vehicle designs for older adults should prioritize efficiency by minimizing demands on learning and memory processes during task execution.
One way to address this accessibility requirement is to delegate some processes to a trusted \mbox{individual}.

\section{Project Context}
\label{sec: projektkontext}
In the UNICAR\textit{agil} project, a total of four concept vehicles for autonomous driving with different characteristics were built. 
One of them is the \textit{auto}ELF, a private family vehicle where all members of the family should be able to travel with the vehicle unaccompanied by each other \cite{Woopen2018, Schraeder2025}.
A major advantage of this vehicle is that it can be used independently by people who previously had to rely on an accompanying person when using a car.
In order to be able to react and design specifically to such users, methods for requirements analysis were developed and carried out in previous work \cite{Graubohm2020, Schraeder2019, Schraeder2021}.
In addition, one goal in the development in the project was to adapt the interior of the respective vehicles to the use case to be achieved \cite{Woopen2018}.
For example, an access aid in the form of a lifting platform was built into the \textit{auto}ELF and the design of the vehicle made it possible to transport a wheelchair. 
And the software integrated into the vehicle was designed with a focus on the user group of persons in need of support.

It is important to note that the \textit{auto}ELF is not a vehicle that is designed for \textit{all} user groups.
Passengers, who require intensive care due to their limitations, were not considered in the design of the vehicle \cite{Schraeder2025}.

\section{Example of the App in the Context of \textit{auto}ELF}
\label{sec: example}

\begin{figure}[ht]
     \centering
     \begin{subfigure}[b]{0.2\textwidth}
         \centering
         \includegraphics[width=\textwidth]{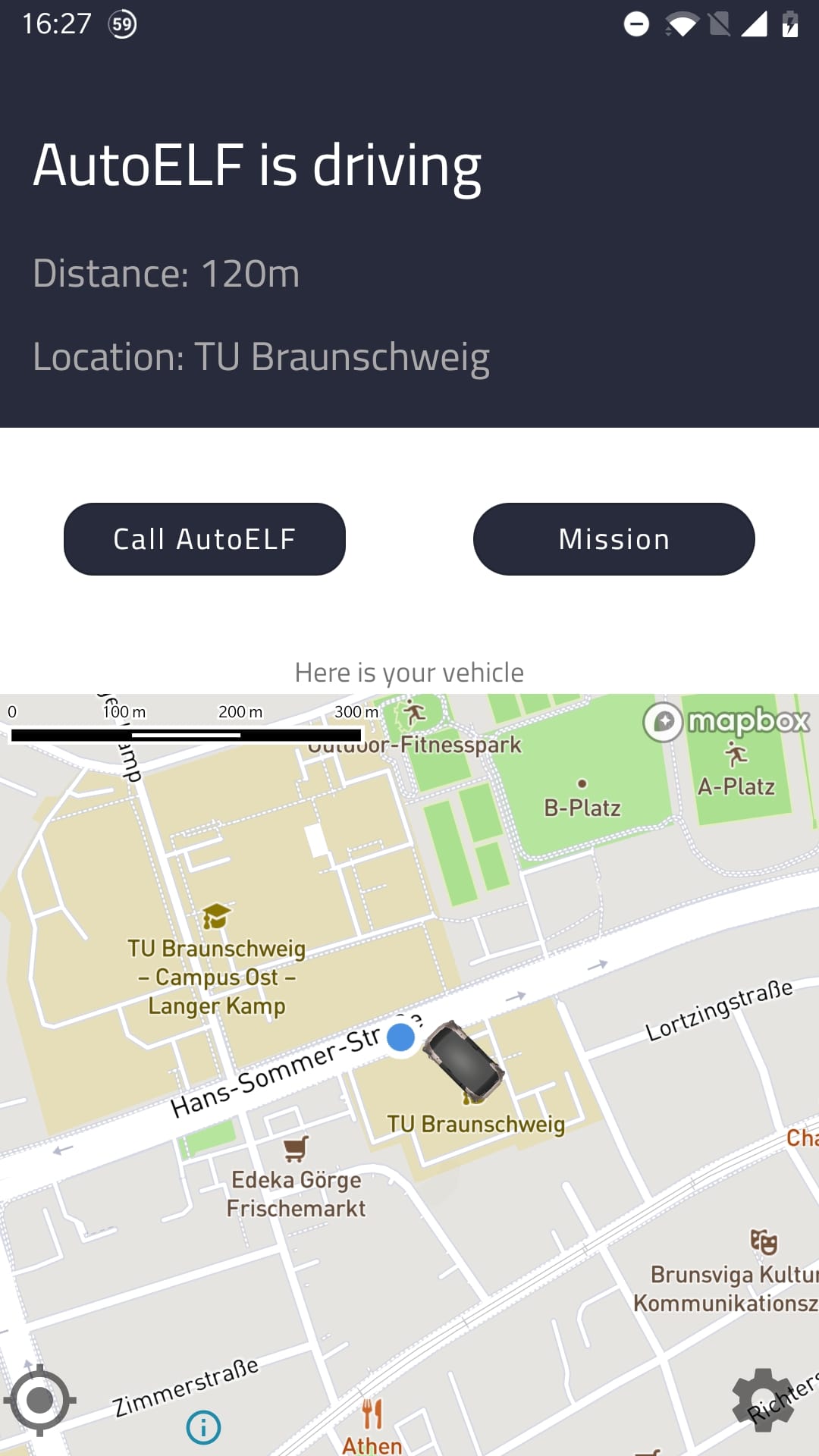}
         \caption{Home Screen}
         \label{fig:app details 5}
     \end{subfigure}
     \hfill
     \begin{subfigure}[b]{0.2\textwidth}
         \centering
         \includegraphics[width=\textwidth]{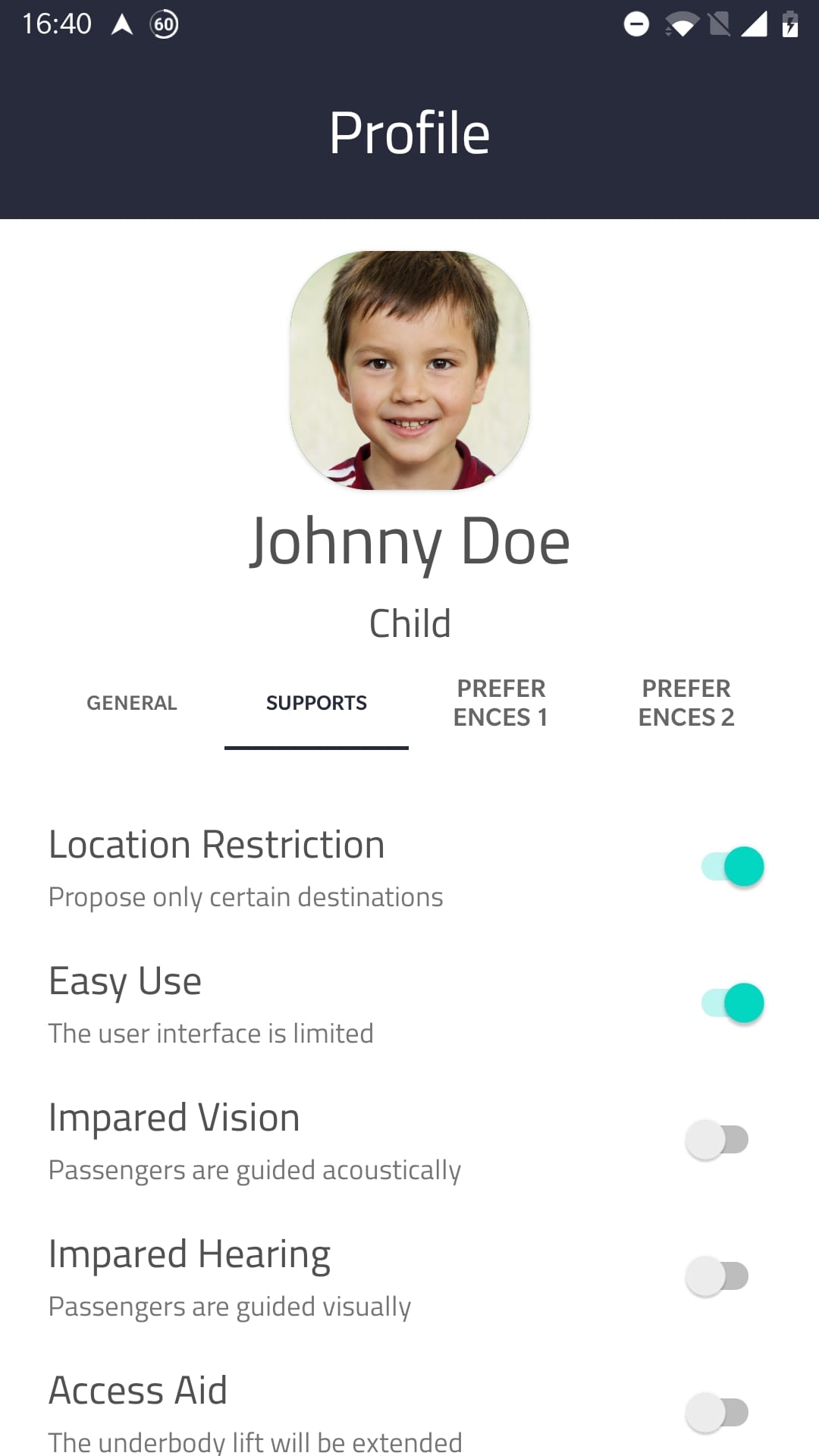}
         \caption{Example settings}
         \label{fig:app details 2}
     \end{subfigure}
        \caption{Selected screens of the app. The screens are selected based on their main functionality to (a) show information about the vehicle and (b) to select settings for the users of the car. 
        The child's face is generated by \url{https://this-person-does-not-exist.com/}}
        \label{fig:app details}
\end{figure}

For the \textit{auto}ELF vehicle, a prototype implementation of the companion app was developed as an Android app in an experimental development.
Meaning, that as no base reference version exists, the app was developed from scratch and is not based on an existing app.
The vehicle itself served as a flexible user platform, which we could design with these specific user groups and their addressed values in mind.

The app is divided into two parts, corresponding to the requirements mentioned in section \ref{sec: requirements}:
The first part contains information about the vehicle: Where it is, on which route it is traveling and a possibility to call the vehicle.
In the second part, it is possible to view and adjust settings for created users: In several subcategories, properties of the vehicle can be selectively specified for the respective users.

Not all of the settings were implemented in the vehicle.     
\textit{auto}ELF does not have a road approval and therefore cannot be used on public roads, which means that not all potential settings could be implemented and were only implemented as dummy functions.

Implemented were, for example, the localization of the vehicle, the light control via the app, the management of user profiles, the retrieval of information from the vehicle and the creation of missions in the app that could be used in the vehicle.  
Not implemented were, for example, the calling of the vehicle to the current position of the user and the execution of the mentioned missions, as the vehicle is not allowed to drive in real road traffic.
In addition, the options for restricted seeing and hearing were not implemented due to the experimental approach in the interior of the \textit{auto}ELF.

Figure \ref{fig:app details 5} shows the home screen of the app.
The current position can be monitored from this screen.
If the car is not currently in use, it can also be called to the current position via this menu.
At the upper part of the app, brief information about the current status of the app is displayed: whether the vehicle is moving, how far the vehicle is from its current position and where the vehicle is located.

In Figure \ref{fig:app details 2}, parts of the setting options are shown that the vehicle has to offer for people in need of support.
Here, the settings of the vehicle can be adapted to the respective person. 

The list below shows some of the settings possible for a companion app.
The focus here is not on the completeness of all possible options, but only to show what is feasible with an app for an autonomous vehicle.
For the selection of the exemplary settings in the table, special attention was paid to settings that could be relevant for people with some kind of disabilities and other people in need of support, such as children and older adults.

\begin{itemize}
    \item Reminder of entry and exit
    \item Preferences for exit (Duration of exit, access aid)
    \item Location restriction
    \item Simplified Operation 
    \item Child safety lock (Disallowing specific vehicle functions)
    \item Exit conditions (e.g., pick-up persons)
    \item Measures if passenger does not appear

    \item Passenger rights settings (access to settings, profile management, mission planning)
    \item Measures in case of interruptions

    \item Comfort preferences                  
    \item Entertainment preferences            
    \item Driving settings (adapt driving behavior for e.g., kinetosis prevention)                    
    \item Control options (e.g., gesture or voice control)
    \item Identification options               
    \item Activate Health/behavioral monitoring       

    \item Locations to drive to/weekly routines
    \item Set Name and Age
    \item People/User in the app to care for
    \item Options to adapt UI for cognitive, motoric or sensory impairment               
\end{itemize}

The list shows that there are quite a few setting options for individual users.
Depending on which person is currently driving in the vehicle, other options may be more important than others.
The list is based on a concept that was developed in collaboration with a bachelor's thesis~\cite{Lampe2022}.

\section{Conclusion}
\label{sec: conclusion}

Autonomous vehicles offer new ways to deal with \mbox{passengers}.
Especially the fact that the ride is no longer limited to people who have are able to drive can lead to new opportunities.
In this contribution, the possibilities of an companion app as a supporting technology to align the capabilities of the vehicle for passenger in need of support were shown. 
With the help of a value-based design, the ride alone can either be enabled or adapted to the respective person.

Such a companion app can address stakeholder values and provide added value through the support, settings and information options offered, especially in combination with persons in need of support.
However, the safety aspects (Safety, Security and Privacy) of this technology must also be highlighted.
The introduction of such technology always carries the possibility of attacks or errors that can lead to dangerous situations, which must be taken into account during development.

In this work, an experimental approach was used to develop a companion app without a base reference model. Positioning it in the define and design phase in a human-centered design approach.
The aim of this contribution is to aid the future development of such an app and provide initial considerations.

This companion app was developed as part of the UNICAR\textit{agil} project and presented together with \textit{auto}ELF in summer 2023.
The app was also part of a user study on the entire \textit{auto}ELF user platform during its development~\cite{Stange2024}.

\section*{Acknowledgment}
The translation of this article was supported by the use of ai translation tools.
The article was first written in German and then translated carefully into English.

We would especially like to thank Daniel Stein, Immo Mädge, Ricco Müller, Jasper Sünnemann, and Henrik Lampe, who contributed as research assistants in the development of the app.

\renewcommand*{\bibfont}{\footnotesize} 
\printbibliography

\end{document}